\documentclass[useAMS,usenatbib]{mn2e}
\usepackage{graphicx}
\usepackage{caption}
\usepackage{xcolor}

\title[Pulsating components in binary systems and the case of DY~Aqr]{OMC/INTEGRAL
photometric observations of pulsating components in eclipsing binaries and
characterization of DY~Aqr}
\author[J. Alfonso-Garz\'{o}n et al.]{J. Alfonso-Garz\'{o}n$^{1}$\thanks{E-mail:
julia@cab.inta-csic.es}, B. Montesinos$^{1}$, A. Moya$^{1}$, J. M. Mas-Hesse$^{1}$ and S.
Mart\'{\i}n-Ruiz$^{2}$\\
$^{1}$Centro de Astrobiolog\'{\i}a -- Departamento de Astrof\'{\i}sica (CSIC-INTA), POB
78, 28691 Villanueva de la Ca\~nada, Spain\\
$^{2}$Instituto de Astrof\'{\i}sica de Andaluc\'{\i}a - CSIC, Glorieta de la
Astronom\'{\i}a s/n, 18008 Granada, Spain}
\begin{document}

\date{Accepted 2014 July 8.  Received 2014 July 8; in original form 2014 January 13}

\pagerange{\pageref{firstpage}--\pageref{lastpage}} \pubyear{2014}

\maketitle

\label{firstpage}

\begin{abstract}
We present the search for eclipsing binaries with a pulsating component in the first catalogue of optically variable sources observed by OMC/INTEGRAL, which contains photometric data for more than 1000 eclipsing binaries. Five objects were found and a detailed analysis of one of them, DY~Aqr, has been performed. Photometric and spectroscopic observations of DY~Aqr were obtained to analyse the binary system and the pulsational characteristics of the primary component. By applying the binary modelling software {\sc phoebe} to the OMC and ground-based photometric light curves, and to the radial velocity curve obtained using echelle high-resolution spectroscopy, the physical parameters of the system have been determined. Frequency analysis of the residual data has been performed using Fourier techniques to identify pulsational frequencies. We have built a grid of theoretical models to classify spectroscopically the primary component as an A7.5V star (plus or minus one spectral subtype). The best orbital fit was obtained for a semi-detached system configuration. According to the binary modelling, the primary component has $T_{\rm eff} = 7625 \pm 125$ K and $\log g = 4.1 \pm 0.1$ and  the secondary component has $T_{\rm eff} = 3800 \pm 200$ K and $\log g = 3.3 \pm 0.1$, although it is too faint to isolate its spectral features. From the analysis of the residuals we have found a main pulsation frequency at 23.37 c/d, which is typical of a $\delta$~Scuti star. In the O-C diagram no evidence of orbital period changes over the last 8 years has been
found.
\end{abstract}


\section{Introduction}

Asteroseismology aims to understand the structure and evolution of stars by examining
their oscillation modes \citep{brown1994, aerts2010, chaplin2013}. Among the classical
pulsators, $\delta$~Scuti stars, which are located at the intersection of the Cepheid
instability strip and the main sequence, where the stellar structure transits from a
highly developed outer convective zone to its absence \citep{rodriguez2001}, are rather
common. They pulsate with amplitudes from milimagnitudes up to tenths of a magnitude
\citep{breger2000}, but only a fraction of the theoretically predicted modes are observed,
which results in many free parameters in the pulsation models. It is therefore important
to infer some of the fundamental stellar properties of these pulsators from other methods,
like for example by studying the pulsating components in binary (multiple) systems.  In
this kind of systems it is possible to characterize precisely their components, supplying
additional constraints for a more reliable modelling. About 70\% of all stars in the Solar
neighbourhood are members of binary or multiple systems ($\sim$ 60-70\% for late-type
stars \citep{mayor2001, chini2013}, $\sim$ 70-80\% for early-type stars \citep{mason2009,
sana2011, chini2012}). This fact is usually ignored in the study of stellar pulsations
although there is strong evidence that binarity might affect the pulsation properties in
specific cases (e.g. the eccentric binaries HD 177863 \citep{decat2002}, HD~209295
\citep{handler2002} and KIC~4544587 \citep{hambleton2013}). It is thus essential to
understand the possible link(s) between binarity and pulsation \citep{lampens2006}. Recent
studies point to the fact that single and binary-member $\delta$~Scuti stars can show
different behaviour because of their different evolutionary status \citep{mkrtichian2003,
soydugan2006, liakos2012}. \citet{mkrtichian2004} introduced the \textit{oscillating
eclipsing Algols} (oEAs) as main sequence pulsators in semi-detached Algol-type eclipsing
binary systems. For several of these systems, they found variable pulsational
characteristics which could be related to mass transfer episodes or tidal distortions.

Several surveys have been carried out in the last years to increase the number of known objects of this type. \citet{soydugan2006} compiled a catalogue of close eclipsing binary systems (detached and semi-detached ones) with one $\delta$~Scuti component.
\citet{zhou2010} made a compilation of 190 oscillating eclipsing binaries (60 of them
containing a $\delta$~Scuti component) and more than 370 pulsating binary or multiple
stellar systems. \citet{liakos2012} published a catalogue with 74 eclipsing binaries with a $\delta$~Scuti component, and this number is increasing at a very rapid rate thanks to the results of \textit{MOST} \citep{walker2003}, \textit{CoRot} \citep{baglin2009} and \textit{Kepler} \citep{borucki2010,gilliland2010}. However, just a few members of this group have been investigated in depth (some recent examples are \citet{lehmann2013, maceroni2014, dasilva2014}).

In this work we present the search for eclipsing binaries with a pulsating component in the first catalogue of variable objects observed by the Optical Monitoring Camera (OMC; \citet{mas-hesse2003}) onboard the INTEGRAL observatory \citep{winkler2003}, OMC--VAR \citep{alfonso-garzon2012}. A detailed follow-up study of the physical properties of DY~Aqr, an Algol-type binary with its primary component showing $\delta$~Scuti oscillations, has been performed, by combining optical photometry obtained with OMC and the 90-cm telescope at the Observatorio de Sierra Nevada (OSN, Granada, Spain), with high-resolution spectra taken with the HERMES instrument on the MERCATOR telescope at the Observatorio del Roque de los Muchachos (ORM, La Palma, Spain)). In Sect. \ref{sec:search}, we show the results of the search. A short introduction to DY~Aqr is included in Sect. \ref{sec:DYAqrintro}. In
Sect.~\ref{sec:obs}, we present the observations obtained to characterize DY~Aqr. The
results of the spectral classification are discussed in Sect. \ref{sec:spclass}. In Sect.~\ref{sec:O_C}, the O-C evolution is analysed. In Sect.~\ref{sec:analysis}, we explain the modelling of the binary features from the photometric and radial velocity curves. Sect.~\ref{sec:pulsation} contains the results of the pulsational frequency analysis performed after subtracting the orbital fit. In Sect.~\ref{sec:evol} we analyse the evolutionary state of the system, considering DY~Aqr is a close binary system. Finally, in \ref{sec:pulcor}, the possible relation between the orbital and pulsational period of DY~Aqr is examined.

\section{The search for eclipsing binaries with pulsating components in OMC--VAR}
\label{sec:search}
\markboth{\ref{sec:search} The search}{Chapter \ref{chap:3}: Pulsating stars
in eclipsing binaries in OMC--VAR }

The Optical Monitoring Camera (OMC) onboard INTEGRAL provides photometry in the Johnson
$V$-band. The first edition of the catalogue of variable objects observed by OMC,
OMC--VAR, contains photometric and variability information for more than 5000 objects,
resulting from the analysis of an initial sample of more than 6000 light curves observed
by OMC during its first 8 years of operations. As discussed in \citet{alfonso-garzon2012},
there are many objects in the OMC--VAR catalogue classified as eclipsing binaries
according to the Variable Star Index (VSX; \citet{watson2006,watson2012}). The VSX is a
database maintained by the American Association of Variable Stars Observers (AAVSO)
containing constantly updated and revised information about variable stars. Out of the
1132 classified eclipsing binaries, we found a period in the literature or have determined
it from OMC data, for 1072 systems. Whenever available, we have used in this work the
values of the periods extracted from our OMC data (see Table \ref{table3_1}). The three types of eclipsing binaries in Table \ref{table3_1}: Algol (EA), $\beta$ Lyrae (EB) and W Ursae Majoris (EW), correspond to the classification in the General Catalogue of Variable
Stars (GCVS; \citet{samus2009}), adopted in the VSX, and is based purely on the shape of
the light curves.

\begin{table}
\caption{Eclipsing binaries in OMC--VAR and period information available.The three types
(EA, EB, EW) correspond to the classification in the GCVS (see text).}
\label{table3_1} 
\begin{center}
\begin{tabular}{l c c c} 
\hline 
Eclipsing type & All & with $P_{\rm OMC}$ & with $P_{\rm VSX}$ \\ 
\hline 
Algol (EA) & 728 & 272 & 703 \\
Beta Lyrae (EB) & 191 & 145 & 184 \\
W Ursae Majoris (EW) & 117 & 100 & 108 \\
Other & 96 & 23 & 61 \\
\hline 
\end{tabular}
\bigskip
\label{tabla3_1}
\end{center}
\end{table}

\begin{figure*}
\includegraphics[width=0.9\textwidth]{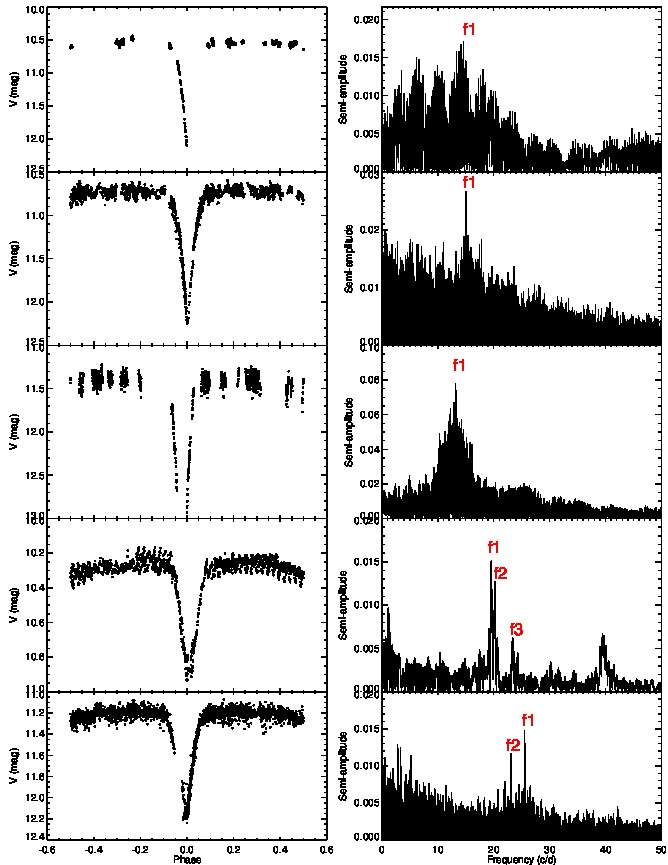}
\caption{Folded OMC photometric light curves and Fourier periodograms from \textsc{period04} of the
five eclipsing binary systems with a $\delta$~Scuti component found in the OMC--VAR
catalogue. From top to bottom the plots correspond to Y~Cam, AW~Vel, MX~Pav, DY~Aqr and
BG~Peg. The frequencies detected for each source are marked in red in the periodograms.}
\label{periodog}
\end{figure*}

In order to detect pulsations in the components of these systems, we have analysed all
eclipsing binaries in OMC--VAR, by modelling their folded OMC light curves with
\emph{polyfit} \citep{pra2008}, aiming to fit just their overall shape. Then we subtracted
the fitted curves from the original data. The residuals obtained, avoiding the intervals
when eclipses occur, were analysed for periodic signals by calculating  the Fourier
periodogram. With this method we confirmed four previously known eclipsing systems with a
pulsating component and we proposed a new candidate, AW~Vel. We have found pulsational
periods similar to the values previously reported in the literature for Y~Cam, MX~Pav and
BG~Peg. For DY~Aqr, the pulsational period measured in previous works is also detected,
but it is not the main one. These results are sumarized in Table \ref{table3_2} and the
folded light curves and periodograms are plotted in Fig. \ref{periodog}.

\begin{table*}
\caption{Results from the search of pulsating components in eclipsing binaries.}
\label{table3_2}
\centering
\begin{tabular}{llcclll}
\hline
Name &  SpT (A+B) & $P_{\rm orb}$ & $P_{\rm puls}~(lit.) $ & Reference & $P_{\rm
puls}~(this~work) $ & Semi-amplitude \\
 &   & (days) & (days)  &  & (days) & (mag) \\
\hline 
Y Cam & A9IV+K1IV & 3.3057 & 0.0665 & 1, 2, 3, 4, 5 & 0.0685(1) & 0.018(2) \\
AW Vel & A4 + [G2IV] & 1.9925 &  &  & 0.0664700(1) & 0.026(1) \\
MX Pav & A5 + K3IV & 5.7308 & 0.0756 & 6, 7 & 0.0756005(2) & 0.081(2) \\
DY Aqr & A0 + [F3]  & 2.1597 & 0.0427 & 8, 9 & 0.051113(1) & 0.0141(7) \\
BG Peg & A2 + [G2IV] & 1.9527 & 0.0391 & 8, 10 & 0.03912730(4) & 0.014(1) \\
%
\hline
%
\end{tabular}
\\
\begin{flushleft}
 (1) \citet{broglia1974}; (2) \citet{broglia1984}; (3) \citet{kim2002}; (4)
\citet{rodriguez2007}; (5) \citet{rodriguez2010}; (6) \citet{michalska2007}; (7)
\citet{pigulski2007}; (8) \citet{soydugan2009}; (9) \citet{soydugan2010}; (10)
\citet{soydugan2011}
\end{flushleft}
\end{table*}

\section{Introduction to DY~Aqr}\label{sec:DYAqrintro}
DY~Aqr was identified as an Algol-type binary showing $\delta$~Scuti oscillations by
\citet{soydugan2009, soydugan2010}. The only information known before this study about the
spectral type of the components was given by \citet{svechnikov1990} and
\citet{malkov2006}, who proposed it to be A0+[F3]. \citet{zasche2011} modelled the binary
system using only the OMC photometric light curve, assuming the primary component to be an
A0~V star. As far as we know, there is no spectroscopic information available for this
source up to now. The lack of precise multiband photometric and spectroscopic data led us
to perform a more detailed follow-up study of this binary system, which has allowed us to
refine the properties of its components.

\section{Observations}\label{sec:obs}

DY~Aqr was observed by OMC for 5 days in two different epochs in 2002 and 2003. With the
aim of characterizing the system, ground-based multicolour photometry and high-resolution
spectroscopy were carried out and analysed. We obtained time for 6 nights of photometric
and 6 nights of spectroscopic observations.


\subsection{Photometric observations}\label{subsec:phot_obs}
OMC photometric data for DY~Aqr in the Johnson $V$ filter have been extracted from the OMC
database \citep{gutierrez2004}. DY Aqr was observed on 24-26 December 2002 and 7-8
November 2003. The OMC light curve has 1699 photometric measurements.

Additional multi-colour photometric observations were collected at the Observatorio de
Sierra Nevada (OSN, Granada, Spain) on 7-8 and 10-13 September 2012. Str\"{o}mgren $uvby$
photometry was carried out using the six channel $uvby\beta$ photometer attached to the
90-cm telescope at OSN (OSN/T90) to determine the physical characteristics of the
components of the system and to perform a detailed study of the pulsational frequency
peaks through different filters. These light curves have 369 photometric measurements each one.


The OSN light curves provide differential magnitudes from a comparison star, HD~211250
($V$ = 8.51 mag). Since additional $UBVJHK$ photometry for HD~211250 is available in
SIMBAD, we have derived the apparent fluxes in the Str\"{o}mgren filters modelling the
Spectral Energy Distribution (SED) of the star and doing synthetic photometry on the
model. Once the Str\"{o}mgren magnitudes for the comparison star are obtained, the
photometry for our target can be directly derived. The best fit for the SED of HD~211250 was obtained
for $T_{\rm eff}=6900$~K, $\log g = 3.5$ and solar metallicity. Another comparison star,
HD~211517, was used to check the stability of the photometry.

\subsection{Spectroscopic observations}\label{subsec:spec_obs}

We obtained time series of high-resolution spectra for 6 nights from 27 August to 2
September 2012. For each observation, we took 3 or 4 exposures of 1200 s, which were
combined in order to reject cosmic rays and to get better signal-to-noise ratio. In this way, 22 combined spectra were obtained in this campaign, covering reasonably well the different phases in the folded light curve. All spectra were taken with HERMES, a fibre-fed prism-cross-dispersed echelle spectrograph on the MERCATOR telescope, in the high-resolution mode (HRF, R$\sim$85000), covering a wavelength range between 3800 and 8750~\AA. The MERCATOR Telescope is a 1.2 m semi-robotic telescope located at the Observatorio del Roque de los Muchachos (ORM) on La Palma (Canary Islands, Spain).

The calibration of the spectra was performed with the automated reduction pipeline
\citep{raskin2011}. It performs the corrections for the bias level, the inter-order
background level, the fringing on the detector, and the modulation of the intensity in
each spectral order and applies a pre-normalisation eliminating the global
wavelength-dependency of the flatfield calibration system. Finally, it applies a
(non-absolute) flux calibration to each order and merges them into a single spectrum.

\section{Spectral classification}\label{sec:spclass}

The spectral type and physical parameters of the primary component of DY~Aqr were
determined using the spectra obtained with HERMES/MERCATOR. 

\begin{figure*}
\includegraphics[width=60.5mm]{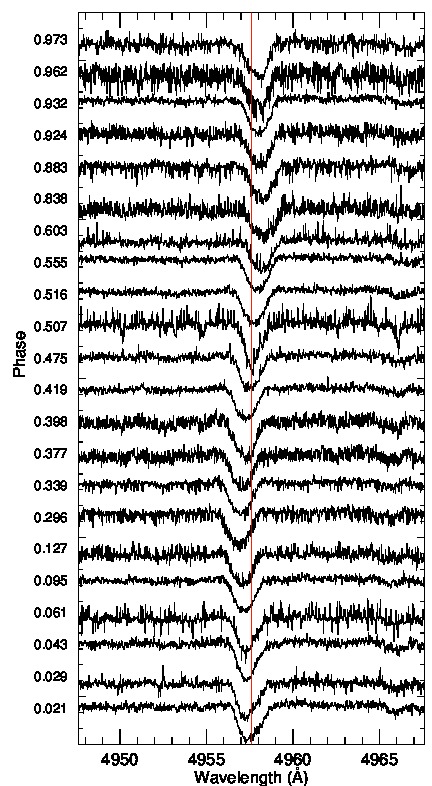}
\includegraphics[width=55mm]{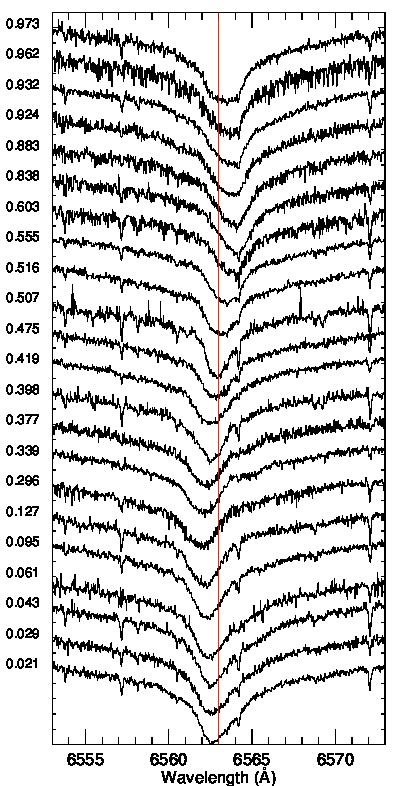}
\includegraphics[width=55mm]{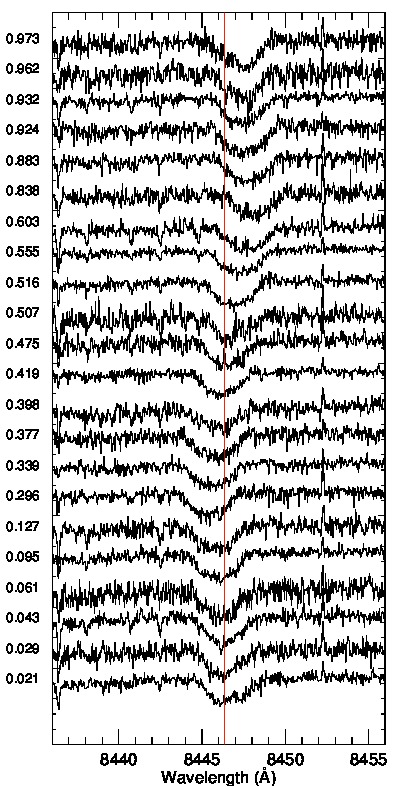}

\caption{The 22 OSN spectra vertically ordered by orbital phase in the regions of Fe\,{\sc i} 4957.60 (left), H$\alpha$ 6562.80 (middle) and O\,{\sc i} 8446.36 (right). Arbitrary units have been used in the y-axis corresponding. The normalized fluxes and the orbital phase is labelled for each spectrum. The solid lines represent the rest wavelengths.}
 \label{fig:spectra}
\end{figure*}

With the aim of separating each component spectra, we applied the spectral disentangling
technique, using {\sc korel}, a {\sc fortran} code for disentangling spectra of binary and multiple systems which uses Fourier transform in the wavelength domain
\citep{hadrava1995, hadrava1997}. {\sc korel} assumes that an observed double-lined
spectrum is a simple linear combination of two single-lined spectra whose velocities
reflect the orbital properties of the system. When applied to a set of spectra covering
the orbital phases, {\sc korel} extracts the contribution of each component and provides
the corresponding velocities relative to the system barycenter. For the 22 spectra, we
have studied different regions along the spectral range, covering the most significant
spectral lines (Ca\,{\sc ii}-K 3933.73~\AA, Fe\,{\sc i} 4957.60~\AA, Na\,{\sc i} D 5889.95~\AA, 5895.92~\AA, H$\alpha$ 6562.80~\AA, O\,{\sc i} 8446.36~\AA~and Ca\,{\sc ii} 8662.14~\AA) and we have found that in all the cases the secondary contribution was below the detection level. In  Fig. \ref{fig:spectra} we have plotted the 22 spectra vertically arranged according to the orbital phase in the regions of Fe\,{\sc i} 4957.60~\AA, H$\alpha$ 6562.80~\AA~and O\,{\sc i} 8446.36~\AA. The solid lines represent the rest wavelengths. No hints of the secondary component can be appreciated. On the other hand, if we examine the profiles of the lines close to the primary minimum, the Rossiter-McLaughlin effect (hereafter RME; \citet{rossiter1924, mclaughlin1924}) can be observed.

Based on the {\sc korel} results and as we will discuss in Sect. \ref{sec:analysis}, we have considered the contribution of the secondary component to the combined spectra to be not significant enough for our purpose of making a spectral classification, since the spectral lines expected to come from it are not detected. Although the contribution to the continuum is about 3-4\% in the V~band (see Table \ref{tab:phoebe_res}), the effect in the spectral lines should be less noticeable. Moreover, the spectral lines are diluted because of the rotational velocity and are probably below the noise level. We have then combined all the available spectra, except those affected by the RME, taking into account the Doppler shifts for each phase point. In this way we obtained a normalized spectrum with higher signal-to-noise ratio to work with.

First of all, we examined the blue region of the spectrum. In this region, the hydrogen lines are enhanced by Stark broadening \citep{mihalas1978}, but these lines become narrower toward later types. For DY~Aqr we measured a width of 33~\AA{} for H$\gamma$ which is compatible with an effective temperature around 7500-8000 K and $\log g$ between 3.5 and 4.5 (see Fig. 5.5 in \citet{gray2009}). Moreover, in this part of the spectrum, there are other features that can be used in the classification of A-type stars, like the comparison of the hydrogen Balmer lines, which are broadest for A2, with the Ca {\sc ii} K-line, which grows in strength as the temperature decreases. Finally, looking at
the near-infrared region of the spectrum, the most interesting trend is the appearance of
the hydrogen Paschen lines. In the early A-type stars, these lines are much stronger than
the near-infrared Ca {\sc ii} triplet, but in the late A-type stars and in F-type stars, the Ca {\sc ii} triplet becomes stronger. As the effective temperature decreases, the P13, P15 and P16
lines become stronger relative to the other unblended Paschen lines. On the other hand,
the O {\sc i} $\lambda$8446 and some N {\sc i} lines in this region are sensitive to
gravity. Examining these lines, the spectrum of DY~Aqr corresponds to a main sequence
star.

According to these criteria, the primary component has to be a main sequence star with an
effective temperature around 7500-8000 K. To quantify its properties, we synthesized a set
of high-resolution spectra, using the suite of programmes SYNTHE \citep{kurucz1993}.
SYNTHE takes as input the list of the atomic and molecular transitions and the spectral
range to be synthesized, together with the corresponding model atmosphere
\citep{castelli2003} for a given temperature, gravity and metallicity, including the
atomic fractions of the different elements. SYNTHE computes the excitation and ionization
populations of neutrals and ions and then produces the final spectrum within the defined
spectral range. A special module of SYNTHE computes intensities at several inclinations in
the atmosphere and uses these, along with the value of $v \sin i$, to reproduce the
spectrum of the star.

\begin{figure}
\includegraphics[width=0.48\textwidth]{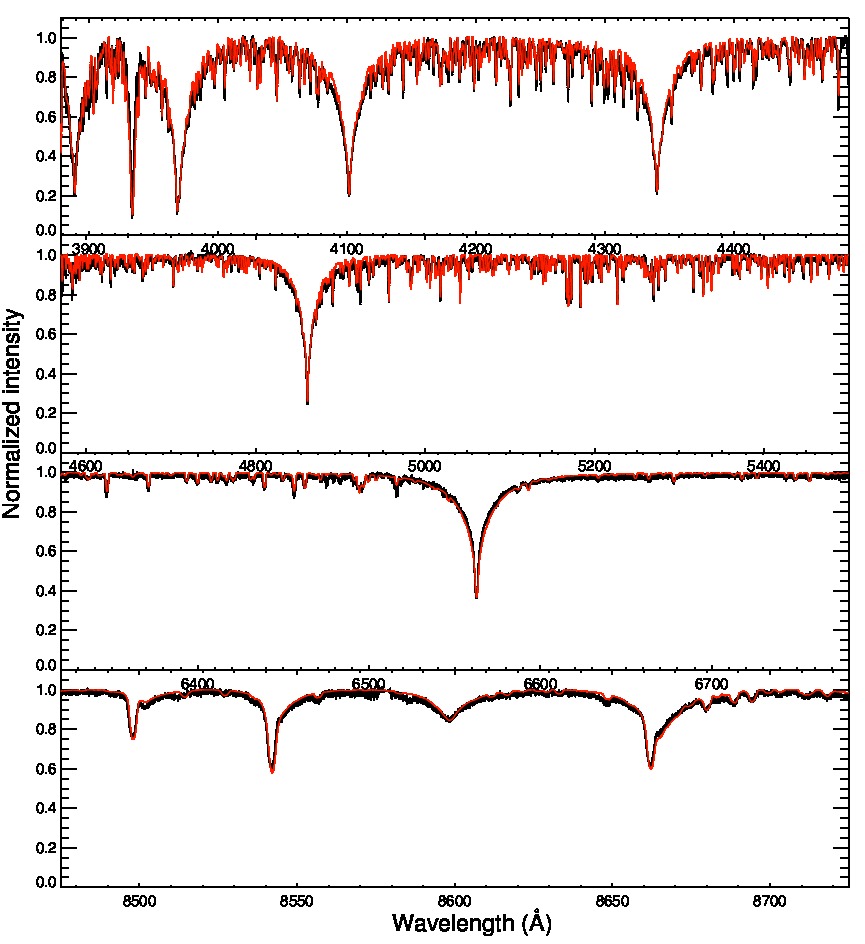}
\caption{Combined spectrum of DY Aqr and the best-fitting synthetic spectrum computed for
$T_{\rm eff}=7625$ K, $\log g_*=4.25$ and solar metallicity (red solid line) broadened
with $v \sin i = 50$ km s$^{-1}$.}
\label{fitspectrum}
\end{figure}

We explored the parameter space with $T_{\rm eff}$ between 7000 and 8500 K and $\log g$
between 3.5 and 4.5, with steps of 125~K and 0.25 dex respectively. The synthetic spectra were broadened with rotational velocities, $v \sin i$, between 0 and 100 km~s$^{-1}$, with a step of 10~km~s$^{-1}$. Typical uncertainties in $T_{\rm eff}$ and $\log g$ are $\pm$125~K and $\pm$0.25~dex, correspondingly. We concluded that the model with $T_{\rm eff}=7625 \pm125$ K, $\log g= 4.25 \pm 0.25$, solar metallicity and $v \sin i = 50 \pm 10$ km~s$^{-1}$ provides the best fit to the data (see Fig.~\ref{fitspectrum}). These parameters correspond to a spectral type A7.5~V \citep{schmidt-kaler1982}, plus or minus one spectral subtype, which indicates that DY~Aqr is much cooler than assumed in the past.

\section{O-C evolution}\label{sec:O_C}

\begin{figure}
\resizebox{\hsize}{!}{\includegraphics{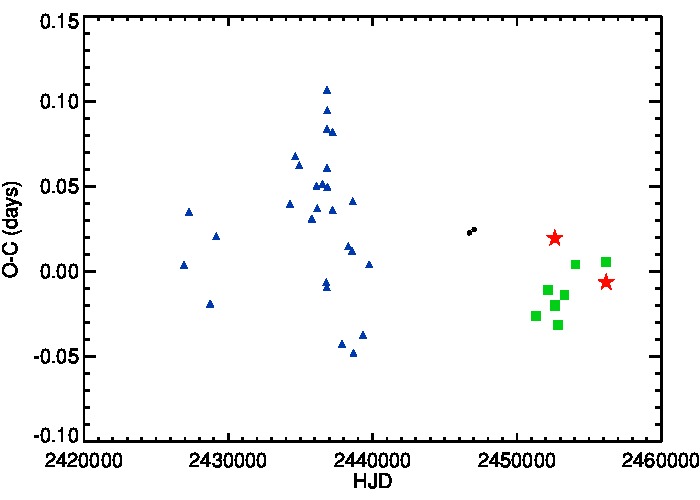}}
\caption{The O-C diagram of DY~Aqr. All the points correspond to primary minima. The small
black dots are from visual observations, the medium blue triangles are from photographic
observations and the bigger green squares are from CCD observations. The red stars
represent the results from this work.}
\label{OmC}
\end{figure}

In order to construct the O-C diagram, the linear ephemerides HJD $= 2426929.485 + \rm E \times 2.159695$ were used \citep{kreiner2004, paschke2006}. Two primary minima at different times were determined from our photometric data, using the method described in \citet{kwee1956}. From the OMC data, we measured a primary minimum at HJD $= 2452634.1946\pm0.0020$ and from OSN data, a minimum was measured at HJD $= 2456182.5476\pm0.0008$. As can be seen in Fig.~\ref{OmC}, no significant long-term changes in the period are appreciated.

\section{Binary modelling}\label{sec:analysis}

It is only after removing the effects from binarity that the pulsational characteristics of the primary component of the system can be analysed on the residual data. The properties of the binary system have been derived from the photometric light curves in five colours (Johnson $V$ from OMC and Str\"{o}mgren $uvby$ from OSN) and from the radial velocity (RV) curve of the primary component extracted from the high-resolution HERMES/MERCATOR spectra.

 
The binary modelling was made using {\sc phoebe} \citep{pra2005}, a package based on the
Wilson-Devinney code (hereafter WD; \citep{wilson1971, wilson1979}). We tried two different configurations, detached and semi-detached with the secondary star filling its Roche lobe, founding that the first one converges to the second one, so we modelled our data with the configuration of a semi-detached system with the secondary component filling its Roche lobe.
First, we analysed separately the radial velocity curve and the multi-colour photometric light
curves and then, we combined all the available data to get the final solution.

\begin{figure}[h]
\includegraphics[width=82mm]{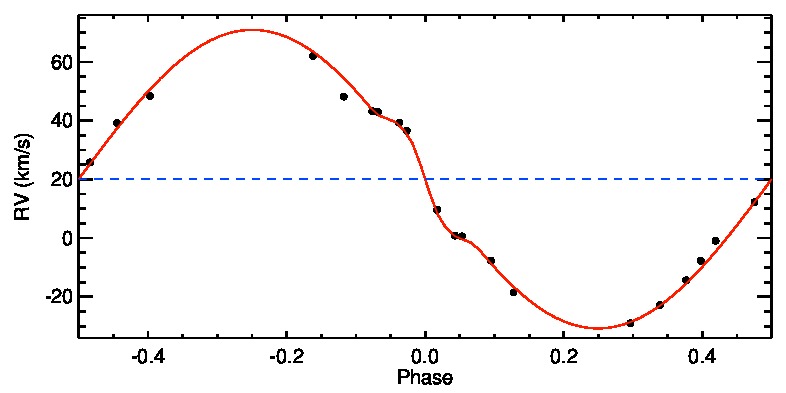}
\caption{Radial velocity curve of the primary component of DY~Aqr. Filled
circles are the \textit{fxcor} measurements. Typical errors are below 1
km~s$^{-1}$. The red solid line is the {\sc phoebe} fit and the blue dashed line
represents the fitted center-of-mass velocity as provided by {\sc phoebe}. Note the clear evidence of the RME around zero phase.}
\label{RV_curve}
\end{figure}

Using the HERMES/MERCATOR spectra we derived the radial velocity curve for the primary
component (see Fig.~\ref{RV_curve}) but not for the secondary, because it was too weak
compared to the primary  to distinguish any of its spectral features. To measure the radial
velocities we used the {\sc iraf} package \textit{fxcor}, with the spectrum closest to the
secondary eclipse (to minimize the secondary component contribution) as template.
According to the spectral type of the star, the best wavelength region of the spectra to
carry out the correlation is 3900-5100~\AA.

When we applied the spectral disentangling technique (see Sect. \ref{sec:spclass}), the
radial velocitiy curves provided by {\sc korel} for the primary component disentangled
spectrum for each region were found to be in good agreement with the results found with
\textit{fxcor}.

In the RV curve, the effects of the RME can be appreciated. The RME occurs in
eclipsing systems when the companion crosses in front of the star; it is due to stellar
rotation and creates a distortion in the radial velocity curve of the star. It strongly
depends on the radii and separation of the stars, on the orbital inclination or impact
parameter and on the position angle of the rotation axis. Its modelling allows to determine these values with a higher precision. In this case, the anomaly is symmetric with a similar redshift during the first half of the eclipse as the blueshift during the
second half, so that we can conclude that the rotational spin and the orbital axis are
aligned. {\sc phoebe} considers the RME and proximity effects in the modelling. In Fig.~\ref{RV_curve}, the RV measurements of the primary (black dots) and the {\sc phoebe} fit of the radial velocity curve (red solid line), as well as the modelled systemic velocity (blue dashed line) are plotted. The RV semi-amplitude of the primary component is $k_{\rm 1} = 49.0 \pm 0.1$ km~s$^{-1}$.

To solve the photometric light curves some basic assumptions were made (see Table
\ref{tabla3_2}). According to the results from the grid of synthetic spectra, the effective temperature of the primary was fixed to 7625~K. We also fixed the orbital period, that was obtained from OMC data and the Phase Dispersion Minimization (PDM) method \citep{Stellingwerf1978}, as
explained in \citet{alfonso-garzon2012}. As the secondary eclipse is centered at phase 0.5 and the light curve appears symmetrical, we have considered that the orbit is circular so the eccentricity $e$ was fixed to 0.

The mass-function has been calculated from the values of the orbital period and the semi-amplitude of the radial velocity curve of the primary component, considering that the orbit is circular. This leads to a value of $0.028\pm0.001$~M$_\odot$. 
The effective temperature and gravity of the primary component, determined from the spectroscopic analysis ($T_{\rm eff} = 7625 \pm 125$ K and $\log g = 4.25 \pm$ 0.25), can be placed on a HR diagram along with the evolutionary tracks from \citet{girardi2002}. This allows to estimate a mass of 1.8~M$_\odot$. The uncertainties in the effective temperature and gravity lead to an uncertainty of $\pm 0.2$ M$_\odot$. As we will see in Sect. \ref{sec:evol}, a classical mass-luminosity relation (MLR) will provide us with an approximate value of the mass, although the luminosity of a star that is a member of a close binary system is usually larger than expected.
From the mass-function and considering the uncertainties in the mass of the primary component, we have estimated the mass of the secondary component to be between 0.51~M$_\odot$ and 0.59~M$_\odot$. The mass-ratio is therefore constrained to the range 0.29 and 0.32, leading to a value of $q = 0.31 \pm 0.02$. Combining this information with the expected radius of the Roche-lobe filling component for a mass-ratio \citep{eggleton1983} and the relative sum of the radii obtained from the width of the eclipses, we could put some constraints in the fit. Theoretical
values of the bolometric albedos (A1 = 1.0 and A2 = 0.6) and gravity-darkening exponents
(g1 = 1.0 and g2 = 0.32) were adopted for the primary and secondary stars, corresponding
to radiative and convective envelopes, respectively, in agreement with their final surface
temperatures \citep{rucinski1969}. The limb darkening coefficients were taken from the
{\sc phoebe} limb darkening tables \citep{pra2011}. The combined depths and widths of the
eclipses were then adjusted by altering the inclination and stellar potentials,
respectively.


\begin{table}
\caption{Fixed parameters for the binary modelling with {\sc phoebe}.}
\label{tabla3_2} 
\begin{center}
\begin{tabular}{lr} 
\hline
Parameter & Values\\
\hline
Time of primary minimum (BJD) &  2452632.03258 \\
Orbital period (d) &  2.15967989(1)\\
Orbital eccentricity, e & 0.0\\
Primary $T_{\rm eff}$ (K) & 7625(125) \\
Primary Bolometric albedo  & 1.0\\
Secondary Bolometric albedo &  0.60\\
Primary gravity brightening  & 1.0\\
Secondary gravity brightening &  0.32\\
Primary rotation  & 1.00\\
Secondary rotation  & 1.00\\
Third light &  0.0\\
\hline
\end{tabular}
\end{center}
\end{table}

\begin{figure}[h]
\includegraphics[width=82mm]{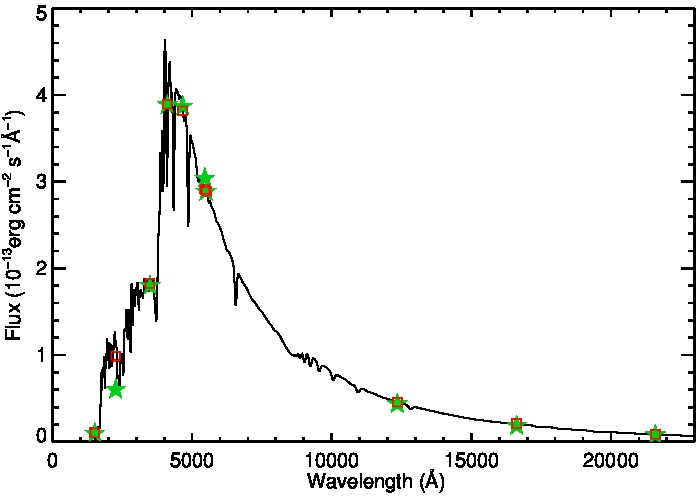}
\caption{Spectral energy distribution of DY Aqr. The green filled stars represent the
observed photometric points taken from GALEX and 2MASS catalogues and our observations.
The black line represents the synthetic composite spectrum, calculated with the parameters
resulting from the binary modelling. The red crosses are the
synthetic photometric points extracted from this spectrum. The synthetic spectrum and
photometric points have been reddened with $E(B-V) = 0.07$ to fit the observations.}
\label{fig:sed}
\end{figure}


With the system parameters resulting from the first iteration of the modelling, we
proceeded to estimate the reddening. We built the spectral energy distribution (SED) of
DY~Aqr, using data from our observations and from the 2MASS \citep{cutri2003} and GALEX
catalogues \citep{bianchi2011}. With the parameters of the first fit, we built the
combined synthetic spectrum and obtained the photometrical points at various bands. By
comparing these photometrical points with the observed ones, we estimated the $E(B-V)$
value corresponding to the derived reddening. The best fit of the models to our SED was
provided assuming a value of $E(B-V) = 0.07$ (see Fig.~\ref{fig:sed}). A standard Galactic
extinction law  \citep{cardelli1989} was used.


Then, the photometric light curves were dereddened using the derived value of
$E(B-V)=0.07$. As the reddening is wavelength dependent we had to calculate the
contribution of the reddening for each filter and for each point of the light curve. In
the case of binary systems, the composite spectrum in each phase point is slightly
different due to geometrical effects and this fact has to be taken into account. The best
way for the treatment of reddening of the light curves is the following: Using the model
spectra of the primary and of the secondary, and the values of the radial velocities of
the primary and the estimated mass-ratio, we can doppler-shift the model spectra and add
them together to get the composite spectrum for each phase point. Then we have to multiply
the composite spectrum by the reddening function. After that, we integrate the resulting
reddened spectrum. If we compare this flux with the one resulting from the same process
but without convolving it with the reddening function, the ratio between the flux obtained
from both cases is the reddening correction for each data point. Doing this for each phase
point and for each filter, the corresponding dereddening corrections were derived. This
process was performed for the five photometric light curves.

\begin{figure}
\includegraphics[width=82mm]{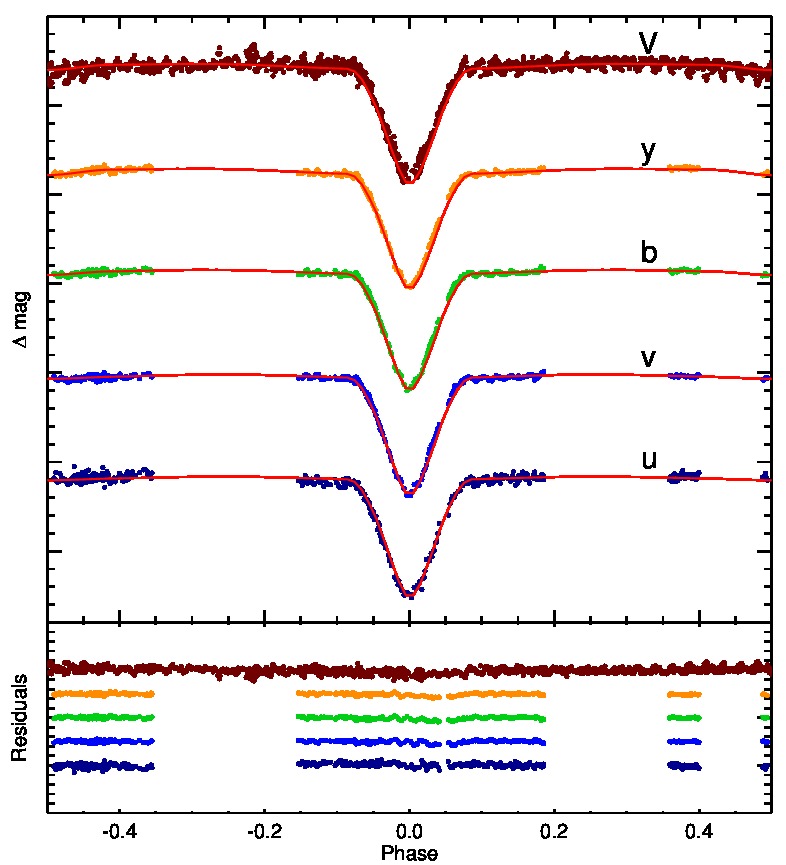}
\caption{\textbf{Top:} Observed and fitted photometric light curves. Solid red lines
represent the synthetic light curves provided by {\sc phoebe} and coloured points
represent the observed photometric light curves for DY~Aqr. \textbf{Bottom:} Residuals of
the fits. In both cases, brown, orange, green, light blue and dark blue points correspond
to V from OMC, Str\"{o}mgren {\it y, b, v} and
{\it u} filters respectively. The light curves have been arbitrarily offset for clarity.}
\label{LCsphoebe}
\end{figure}

Once the light curves were corrected for extinction, a second iteration of the orbital fit
with similar considerations as for the first one was carried out. The final results of the
binary model are summarized in Table \ref{tab:phoebe_res}. Synthetic and observed light curves
are presented in Fig. \ref{LCsphoebe}. The errors assigned to the masses and to the mass-ratio come from the errors of the effective temperature and gravity, as we explained above and the errors of the other parameters were estimated measuring the variations of each value considering the range of different plausible solutions. These error bars are larger than those obtained from the covariation matrix, but are more realistic because many of the parameters involved in the modelling of binary stars are intercorrelated.

\begin{table}
\caption{Parameters and coefficients from the {\sc phoebe} best fit model.}
\begin{center}
\begin{tabular}{lr} 
\hline 
Parameter & Values\\
\hline
\multicolumn{2}{c}{System parameters}\\
\hline
Semi-major axis (R$_\odot$), $a$  & 9.4(5)\\
Mass ratio, $q$ & 0.31(2) \\
Center-of-mass velocity (km $s^{-1}$), $\gamma$ & 20.10(5) \\
Orbital inclination (degrees), $i$ & 75.4(5) \\
Argument of periastron (rad), $\omega$ & 1.57(1)\\
\hline
\multicolumn{2}{c}{Stellar parameters}\\
\hline
Secondary effective temperature (K), $T_{\rm eff 2}$ & 3800(200)\\
Primary mass (M$_\odot$), $M_{\rm 1}$ & 1.8(2) \\
Secondary mass (M$_\odot$), $M_{\rm 2}$ & 0.55(4) \\
Primary radius (R$_\odot$), $R_{\rm 1}$ & 2.1(1) \\
Secondary radius (R$_\odot$), $R_{\rm 2}$ & 2.7(1)\\
Primary potential, $\Omega_{\rm 1}$ & 4.8(1)\\
Primary $\log g$ (cgs), $\log g_{\rm 1}$  & 4.1(1) \\
Secondary $\log g$ (cgs), $\log g_{\rm 2}$  & 3.3(1)\\
Luminosity ratio in $u$ band, L$_{\rm 2}^u/$L$_{\rm 1}^u$ & 0.003(2)\\
Luminosity ratio in $v$ band, L$_{\rm 2}^v/$L$_{\rm 1}^v$ & 0.008(5)\\
Luminosity ratio in $b$ band, L$_{\rm 2}^b/$L$_{\rm 1}^b$& 0.02(1)\\
Luminosity ratio in $y$ band, L$_{\rm 2}^y/$L$_{\rm 1}^y$ & 0.04(2)\\
Luminosity ratio in V band, L$_{\rm 2}^V/$L$_{\rm 1}^V$ & 0.037(1)\\
\hline
\end{tabular}
\label{tab:phoebe_res}
\end{center}

\footnotesize{Note: The values in parentheses give the 1$\sigma$ uncertainty in the
previous digit.}

\end{table}
 
From the derived parameters of the binary modelling and assuming synchronous rotation, we have calculated the projected rotational velocities of each component, obtaining $v_{\rm 1} \sin i = 52 \pm 2$ km~s$^{-1}$ for the primary and $v_{\rm 2} \sin i = 62 \pm 2$ km~s$^{-1}$ for the secondary. The projected rotational velocity of the primary is in good agreement with the measurement from the grid of synthetic spectra. On the other hand, the $v \sin i$ of the secondary is quite high and the implied dilution could explain why we have not detected the spectral lines of the secondary star in our spectra.



\section{Pulsation frequencies}\label{sec:pulsation}

\begin{table}
\caption{Results from frequency analysis of the OSN data.}
\begin{center}
\begin{tabular}{c c c c c} 
\hline
Filter & Frequency & Semi-amplitude & Phase & S/N \\ 
 &  (c/d) & (mag) &  &  \\ 
\hline
$u$ & 23.36(2) & 0.0062 & 0.89 & 4.7 \\
$v$ & 23.37(2) & 0.0058 & 0.52 & 7.6 \\
$b$ & 23.37(2) & 0.0059 & 0.11 & 5.4 \\
$y$ & 23.36(2) & 0.0047 & 0.56 & 6.2 \\
\hline
\end{tabular}
\label{tabla3_4}
\end{center}
\end{table}

After removing the binary model and avoiding the eclipses areas, we derived the frequency
spectrum of the residual data to study the pulsations from the primary component
of DY~Aqr using \textsc{period04} \citep{lenz2004}. As mentioned above, DY~Aqr was
identified as an eclipsing binary system containing a $\delta$-Scuti by
\citet{soydugan2009}. In that work a frecuency of 23.39 c/d and an semi-amplitude of 0.006 mag were measured.

\begin{figure}[h]
\includegraphics[width=82mm]{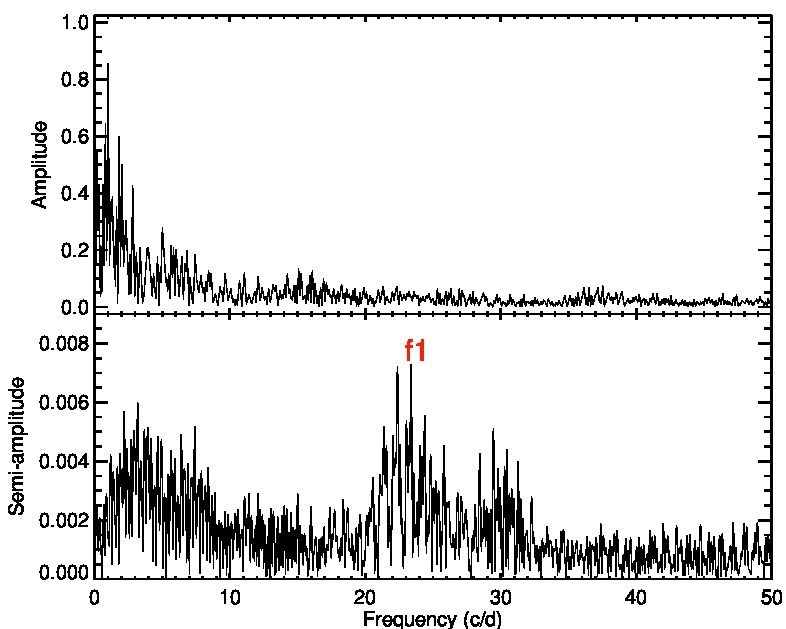}
\caption{Spectral window (top) and power spectrum (bottom) of DY~Aqr, obtained from the Str\"{o}mgren multi-colour photometry acquired at OSN.}
\label{fourier}
\end{figure}

With OMC data the pulsational frecuency of 23.39 c/d found by \citet{soydugan2009} is
confirmed with a semi-amplitude of 0.005 mag, in agreement with the value they measured. However, with OMC data, two more frequencies at 19.56 c/d and
semi-amplitude 0.014 mag and at 20.19 c/d and semi-amplitude 0.013 mag were found, being
more significant than the previously reported one. This could be an indication of mode
amplitude variations with time. In order to check if this was a real change in the
pulsation characteristics of DY~Aqr, we analised separately the Str\"{o}mgren multi-colour
photometry. The frequencies found at each filter are summarized in Table~\ref{tabla3_4}.
We considered a peak to be significant if its S/N ratio is larger than 4.0. The error of
these frequencies is the corresponding Rayleigh frequency resolution $\Delta f =
\frac{1}{\Delta t}$ = 0.16 c/d. The spectral window and power spectrum for the first
frequency are shown in Fig.~\ref{fourier}, respectively. To confirm these results,
we also performed the frequency analysis with \textsc{SigSpec} \citep{reegen2007} and the
agreement between the two methods was good. We did not find any evidence from the analysis
of the OSN data of the two additional frequencies detected in the OMC lightcurve. 

If we compare our results with the values reported by \citet{soydugan2009}, we conclude
that no obvious amplitude variation is detected for the frequency of 23.39 c/d. We have analysed any potential instrumental effects that could explain the other frequencies detected in the OMC data, but none of them is able to justify these frequencies. Moreover, they appear in two different epochs, separated by one year. We plan to continue monitoring DY~Aqr in the future looking for potential short-lived instabilities affecting its pulsation modes.  

\begin{figure*}
\includegraphics[width=75mm]{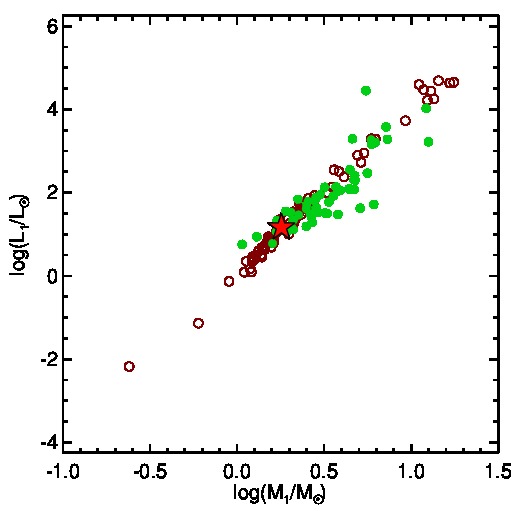}
\includegraphics[width=75mm]{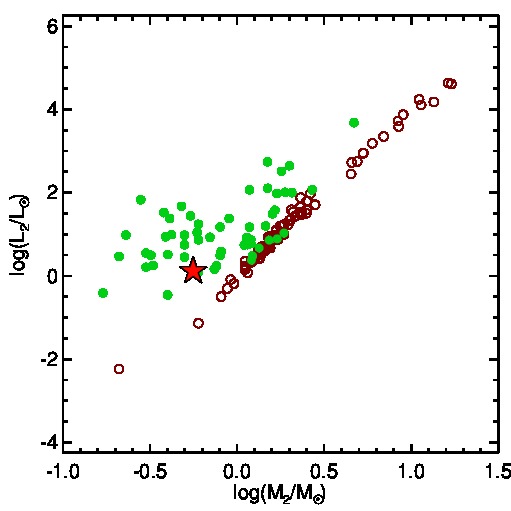}
\caption{Plots adapted from \citet{ibanoglu2006}. Luminosities of the primary components (left) and secondary components (right) of the sample of Algol-type systems compiled by \citet{ibanoglu2006}. Brown open circles represent the detached systems and green circles correspond to semi-detached systems. The red star corresponds to the values for the primary and secondary components of DY~Aqr obtained in this work.}
\label{fig:ibanoglu}
\end{figure*}

\section{Evolution}\label{sec:evol}

It is known that stars in close binary systems evolve in a different way than isolated stars do. Several mass-luminosity relations (MLRs) for eclipsing binaries have been obtained in the last years \citep{gorda1998, malkov2003}. One of the main conclusions of these studies is that the systems with spectral type A and F have higher luminosities (radii) and temperatures than single stars with the same spectral type \citep{malkov2003}. This fact is in good agreement with the results we obtained in our binary modelling. The primary component of DY~Aqr has a larger radius than it would be expected for a single star with its temperature \citep{girardi2002}. \citet{ibanoglu2006} compiled a sample of Algol-type systems with well-determined absolute parameters and found a MLR for the primary and secondary components of detached and semi-detached Algol-type stars. In Fig. \ref{fig:ibanoglu} we have plotted the luminosities of both components of detached and semi-detached systems against their masses as in \citet{ibanoglu2006}. We have overplotted the corresponding values for the primary and secondary components of DY~Aqr obtained in this work. According to these plots, especially because of the behaviour of the secondary component, DY~Aqr seems to present luminosities in agreement with being a semi-detached system.

\begin{figure}
\begin{center}
\includegraphics[width=82mm]{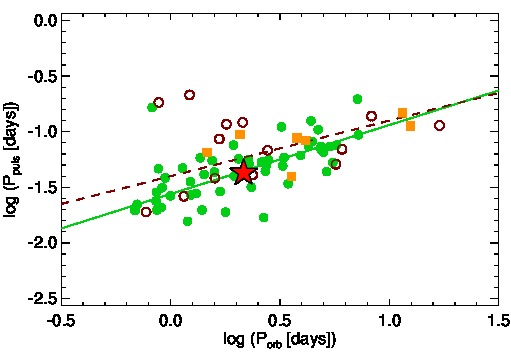}
\includegraphics[width=82mm]{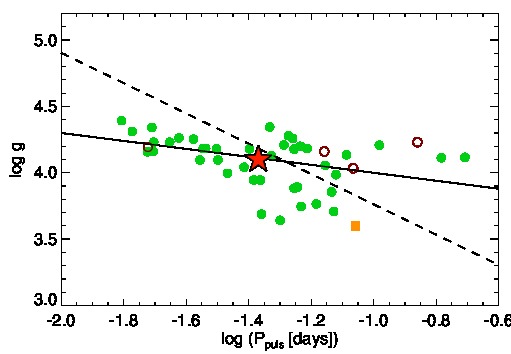}
\includegraphics[width=82mm]{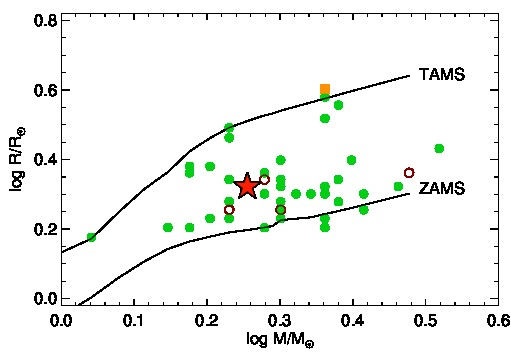}
\end{center}
\caption{In all panels, brown empty circles
correspond to detached systems, green filled circles are semi-detached systems, orange filled
squares are undefined binary systems and the red star corresponds to the pulsating component in the binary system DY~Aqr. {\bf Top:} Correlation between pulsational and orbital periods for $\delta$~Scuti stars in binary systems \citep{liakos2012}. Orange dashed line and green solid line
represent the best fit for detached and semi-detached binaries respectively. {\bf Middle:}
Gravity acceleration versus the dominant pulsation frequency. The dashed line represents
the best linear fit for single $\delta$~Scuti stars derived by \citet{fernie1995} and the
solid line represents the linear fit for $\delta$~Scuti stars members of binary systems
found by \citet{liakos2012}. {\bf Bottom:} M-R diagram for the $\delta$~Scuti stars in
binary systems for the objets in the same survey. }
\label{fig:liakos}
\end{figure}

\section{Correlation between orbital and pulsational period}\label{sec:pulcor}

\citet{soydugan2006} and \citet{liakos2012} discovered a connection between the orbital
and pulsation periods of the oEAs. \citet{liakos2012} found an empirical relation between
the pulsational frequency of the $\delta$~Scuti and the evolutionary state of the system,
differing from the evolutionary behaviour of single $\delta$~Scuti stars.

In Fig.~\ref{fig:liakos} the parameters of DY~Aqr are represented in similar plots as
those shown by \citet{liakos2012}. It can be seen in the figure that DY~Aqr behaves indeed as other semi-detached systems, which is consistent with our results, althought the difference between detached and semi-detached behaviour is not very evident in these plots.
Furthermore, the bottom panel of this figure shows that DY~Aqr is not very evolved yet,
being located close to the ZAMS in the diagram. 

On the other hand, \citet{zhang2013} derived the following theoretical relation between the pulsation and orbital periods of pulsating stars in close binaries, based on their Roche lobe filling:

\begin{equation}
 \log P_{\rm pul} = \log P_{\rm orb} + \log \alpha  
\end{equation}
where
\begin{equation}
 \alpha = \frac{Q}{0.116} f_{1}^{3/2} r_{\rm cr1}^{3/2} (1+q)^{1/2}
\end{equation}
being the pulsation constant $Q = P_{\rm pul} (\rho_{\rm 1}/\rho_{\rm \odot})^{1/2}$, $f_{1} = r_{\rm 1}/r_{\rm cr1}$ the filling factor, $r_{\rm cr1}$ the equivalent radius of the Roche lobe calculated from \cite{eggleton1983}, and $q$ the mass ratio.

Fitting this equation to all the data they collected, they obtained the numerical
relation:
\begin{equation}
 \log P_{\rm pul} = \log P_{\rm orb} -1.70(\pm0.16)
\label{eq:alfa}
\end{equation}

which is very similar to the fit derived by \citet{liakos2012} for semi-detached systems, as shown in the top panel of Fig. \ref{fig:liakos}.

Applying the results obtained in this work, we derived  for DY~Aqr that $r_{\rm cr1} = 0.48$, $f_{\rm 1} = 0.47$, and Q = 0.019~d. This value of the pulsation constant indicates that DY~Aqr pulsates in p-modes, which is expected for a $\delta$~Scuti. If we use the orbital and pulsational periods of DY~Aqr, we get $\log \alpha = 1.703$, which is in perfect agreement with equation \ref{eq:alfa}.

\section{Conclusions}

We have searched for $\delta$~Scuti stars in eclipsing binaries in the OMC--VAR catalogue
of variable objects, confirming four previously detected eclipsing systems with a
$\delta$~Scuti component and proposing a new candidate, AW~Vel. We have performed a
detailed analysis of one of the systems, DY~Aqr, poorly studied in the past, and for which
we
had identified with OMC a possible change in the fundamental pulsational frequencies.
Episodic mass-transfer events in these objects can indeed affect their pulsational
characteristics. 

High-resolution spectroscopy with HERMES/MERCATOR classifies the primary component of
DY~Aqr as an A7.5~V star, much cooler than previously reported.

The O-C diagram does not show any evidence of orbital period changes.

Binary modelling with {\sc phoebe} combining the multicolour photometric ligth curves (OMC and OSN) and the radial
velocity curve of the primary component (HERMES/MERCATOR) constraints the primary
component to an effective temperature of $7625 \pm 125$~K and $\log g = 4.1 \pm 0.1$ and the secondary component to an effective temperature of $3800 \pm 200$~K and $\log g  = 3.3 \pm 0.1$. The detection of the spectrum of the secondary component would provide a better determination of the physical parameters of each component, but the secondary contribution was below the limit of detection in our data.

After removing the binarity effect from the light curves, the pulsational frequencies were
studied through Fourier analysis. We did not confirm with the OSN data the first and
second frequencies detected by OMC at 19.56 and 20.19 c/d. On the other hand, the same
fundamental frequency found by \citet{soydugan2009} at 23.39 c/d, also detected by OMC,
was present in this ground-based photometry. The detection of these 2 additional
frequencies by OMC can not be attributed to known instrumental effects, and point to a
drastic, short-lived change in the pulsational characteristics. A further monitoring would
be required to confirm short lived pulsational instabilities in this star.

The physical parameters obtained for each component of DY~Aqr in the binary modelling are in agreement with the expected evolution of the components of a semi-detached Algol-system as given by \citet{ibanoglu2006}.

Finally, our conclusion of DY~Aqr being an oEA is compatible with the results by \citet{liakos2012} and \citet{zhang2013} on how the mass-transfer in these systems affects their pulsational behaviour, but the detection of the secondary spectrum will be necessary to accurately determine the physical parameters of the system.

\section*{Acknowledgements}

We want to thank the anonymous referee for his/her very helpful comments and suggestions. Thanks to Jes\'{u}s Maldonado for his help with {\sc fxcor}, to Andrej Prs\u{a} and
Kelly Hambleton for their help with {\sc phoebe} and to Petr Hadrava for his help with {\sc korel}. The development, operation and exploitation of OMC have been funded by Spanish MICINN
under grants AYA2012-38897-C02-01 and previous ones.  INTEGRAL is an ESA project funded by
ESA member states (especially the PI countries: Denmark, France, Germany, Italy, Spain,
Switzerland), Czech Republic, Poland, and with the participation of Russia and the USA.
OMC was also funded by Ireland, United Kingdom, Belgium and the Czech Republic. This
research has made use of data from the OMC Archive at CAB (INTA-CSIC), pre-processed by
ISDC, and of the SIMBAD database, operated at CDS, Strasbourg, France.

\bsp

\label{lastpage}


\begin{thebibliography}{99}

\bibitem[{Aerts {et~al.}(2010)Aerts, {Christensen-Dalsgaard}, \&
  Kurtz}]{aerts2010}
Aerts, C., {Christensen-Dalsgaard}, J., \& Kurtz, D.~W. 2010, Asteroseismology
  (Springer {Science+Business} Media {B.V})

\bibitem[{{Alfonso-Garz\'{o}n} {et~al.}(2012){Alfonso-Garz\'{o}n}, Domingo,
  {Mas-Hesse}, \& Gim\'{e}nez}]{alfonso-garzon2012}
{Alfonso-Garz\'{o}n}, J., Domingo, A., {Mas-Hesse}, J.~M., \& Gim\'{e}nez, A.
  2012, A\&A, 548, 79

\bibitem[{Baglin {et~al.}(2009)Baglin, Auvergne, Barge, Deleuil, Michel, \&
  Team}]{baglin2009}
Baglin, A., Auvergne, M., Barge, P., {et~al.}\ 2009, IAU Symposium, 253, 71

\bibitem[{Baker \& Kippenhahn(1961)}]{baker1961}
Baker, N. \& Kippenhahn, R. 1961, AJ, 66, 278

\bibitem[{Bianchi {et~al.}(2011)Bianchi, Herald, Efremova, Girardi, Zabot,
  Marigo, Conti, \& Shiao}]{bianchi2011}
Bianchi, L., Herald, J., Efremova, B., {et~al.} 2011, Ap\&SS, 335, 161

\bibitem[{Borucki {et~al.}(2010)Borucki, Koch, Basri, Batalha, Brown, Caldwell,
  Caldwell, {Christensen-Dalsgaard}, Cochran, {DeVore}, Dunham, Dupree,
  Gautier, Geary, Gilliland, Gould, Howell, Jenkins, Kondo, Latham, Marcy,
  Meibom, Kjeldsen, Lissauer, Monet, Morrison, Sasselov, Tarter, Boss,
  Brownlee, Owen, Buzasi, Charbonneau, Doyle, Fortney, Ford, Holman, Seager,
  Steffen, Welsh, Rowe, Anderson, Buchhave, Ciardi, Walkowicz, Sherry, Horch,
  Isaacson, Everett, Fischer, Torres, Johnson, Endl, {MacQueen}, Bryson,
  Dotson, Haas, Kolodziejczak, Van~Cleve, Chandrasekaran, Twicken, Quintana,
  Clarke, Allen, Li, Wu, Tenenbaum, Verner, Bruhweiler, Barnes, \&
  Prsa}]{borucki2010}
Borucki, W.~J., Koch, D., Basri, G., {et~al.} 2010, Science, 327, 977

\bibitem[{Breger(2000)}]{breger2000}
Breger, M. 2000, in Breger M., Montgomery M. H., eds, ASP Conf. Ser, Vol. 210, Delta Scuti and Related Stars. Astron. Soc. Pac., San Francisco, p. 3

\bibitem[{Broglia \& Conconi(1984)}]{broglia1984}
Broglia, P. \& Conconi, P. 1984, A\&A, 138, 443

\bibitem[{Broglia \& Marin(1974)}]{broglia1974}
Broglia, P. \& Marin, F. 1974, A\&A, 34, 89

\bibitem[{Brown \& Gilliland(1994)}]{brown1994}
Brown, T.~M. \& Gilliland, R.~L. 1994, ARA\&A, 32, 37

\bibitem[{Buzasi {et~al.}(2005)Buzasi, Bruntt, Bedding, Retter, Kjeldsen,
  Preston, Mandeville, Suarez, Catanzarite, Conrow, \& Laher}]{buzasi2005}
Buzasi, D.~L., Bruntt, H., Bedding, T.~R., {et~al.} 2005, ApJ, 619, 1072

\bibitem[{Cardelli {et~al.}(1989)Cardelli, Clayton, \& Mathis}]{cardelli1989}
Cardelli, J.~A., Clayton, G.~C., \& Mathis, J.~S. 1989, ApJ, 345, 245

\bibitem[{Castelli \& Kurucz(2003)}]{castelli2003}
Castelli, F. \& Kurucz, R.~L. 2003, in N. Piskunov, W.W. Weiss, and D.F. Gray, eds, Proc. IAU Symp. 210, Modelling of Stellar Atmospheres. Astron. Soc. Pac., p. A20

\bibitem[{Chaplin \& Miglio(2013)}]{chaplin2013}
Chaplin, W.~J. \& Miglio, A. 2013, ARA\&A, 51, 353

\bibitem[{Chini et~al.(2012)}]{chini2012} 
Chini, R., Hoffmeister, V.~H., Nasseri, A., Stahl, O., \& Zinnecker, H. 2012, MNRAS, 424,
1925 

\bibitem[Chini et al.(2013)]{chini2013} 
Chini, R., Nasseri, A., Dembsky, T., et al. 2013, EAS Publications Series, 64, 155 

\bibitem[{Cutri {et~al.}(2003)Cutri, Skrutskie, van Dyk, Beichman, Carpenter,
  Chester, Cambresy, Evans, Fowler, Gizis, Howard, Huchra, Jarrett, Kopan,
  Kirkpatrick, Light, Marsh, {McCallon}, Schneider, Stiening, Sykes, Weinberg,
  Wheaton, Wheelock, \& Zacarias}]{cutri2003}
Cutri, R.~M., Skrutskie, M.~F., van Dyk, S., {et~al.} 2003, {VizieR} Online
  Data Catalog, 2246, 0

\bibitem[\protect\citeauthoryear{da Silva et 
al.}{2014}]{dasilva2014} da Silva R., Maceroni C., Gandolfi D., Lehmann H., Hatzes A.~P., 2014, A\&A, 565, A55 

  
\bibitem[{De~Cat \& Aerts(2002)}]{decat2002}
De~Cat, P. \& Aerts, C. 2002, A\&A, 393, 965

\bibitem[Eggleton(1983)]{eggleton1983} 
Eggleton, P.~P. 1983, ApJ, 268, 368

\bibitem[{Fernie(1995)}]{fernie1995}
Fernie, J.~D. 1995, AJ, 110, 2361

\bibitem[{Gilliland {et~al.}(2010)Gilliland, Brown, {Christensen-Dalsgaard},
  Kjeldsen, Aerts, Appourchaux, Basu, Bedding, Chaplin, Cunha, De~Cat,
  De~Ridder, Guzik, Handler, Kawaler, Kiss, Kolenberg, Kurtz, Metcalfe,
  Monteiro, Szab\'{o}, Arentoft, Balona, Debosscher, Elsworth, Quirion, Stello,
  Su\'{a}rez, Borucki, Jenkins, Koch, Kondo, Latham, Rowe, \&
  Steffen}]{gilliland2010}
Gilliland, R.~L., Brown, T.~M., {Christensen-Dalsgaard}, J., {et~al.}\ 2010,
  PASP, 122, 131

\bibitem[{Girardi} {et~al.}(2002)]{girardi2002}
Girardi, L., Bertelli, G., Bressan, A., et al.\ 2002, A\&A, 391, 195

\bibitem[\protect\citeauthoryear{Gorda 
\& Svechnikov}{1998}]{gorda1998} Gorda S.~Y., Svechnikov M.~A., 1998, ARep, 42, 793

\bibitem[{Gray \& Corbally(2009)}]{gray2009}
Gray, R.~O. \& Corbally, J., .~C. 2009, Stellar Spectral Classification, Princeton Univ. Press, Princeton, NJ

\bibitem[{Guti{\'{e}}rrez {et~al.}(2004)Guti{\'{e}}rrez, Solano, Domingo, \&
  Garc{\'{i}}a}]{gutierrez2004}
Guti{\'{e}}rrez, R., Solano, E., Domingo, A., \& Garc{\'{i}}a, J.\ 2004, Astronomical
  Data Analysis Software and Systems (ADASS) XIII, 314, 153
  
\bibitem[Hadrava(1995)]{hadrava1995} Hadrava, P.\ 1995, A\&AS, 114, 393

\bibitem[Hadrava(1997)]{hadrava1997} Hadrava, P.\ 1997, A\&AS, 122, 581
  
\bibitem[{Hambleton {et~al.}(2013)}]{hambleton2013}
Hambleton, K.~M., Kurtz, D.~W., Pr\v{s}a, A., {et~al.}\ 2013, MNRAS, 434, 925

\bibitem[{Handler {et~al.}(2002)Handler, Balona, Shobbrook, Koen, Bruch,
  {Romero-Colmenero}, Pamyatnykh, Willems, Eyer, James, \& Maas}]{handler2002}
Handler, G., Balona, L.~A., Shobbrook, R.~R., {et~al.}\ 2002, MNRAS, 333, 262

\bibitem[\protect\citeauthoryear{Ibano{\v g}lu et al.}{2006}]{ibanoglu2006} Ibano{\v g}lu C., Soydugan F., Soydugan E., Dervi{\c s}o{\v g}lu A., 2006, MNRAS, 373, 435 

\bibitem[{Kim {et~al.}(2002)Kim, Lee, Youn, Kwon, \& Kim}]{kim2002}
Kim, S., Lee, J.~W., Youn, J., Kwon, S., \& Kim, C.\ 2002, A\&A, 391, 213

\bibitem[{Kreiner(2004)}]{kreiner2004}
Kreiner, J.~M.\ 2004, Acta Astron., 54, 207

\bibitem[{Kurucz(1993)}]{kurucz1993}
Kurucz, R.~L.\ 1993, {SYNTHE} spectrum synthesis programs and line data:
  Smithsonian Astrophysical Observatory, 1993, 18

\bibitem[{Kwee \& van Woerden(1956)}]{kwee1956}
Kwee, K.~K. \& van Woerden, H.\ 1956, Bulletin of the Astronomical Institutes of
  the Netherlands, 12, 327

\bibitem[{Lampens(2006)}]{lampens2006}
Lampens, P.\ 2006, in Sterken, C. and Aerts, C., eds, ASP Conf. Ser. Vol 349, Astrophysics of Variable Stars. Astron. Soc. Pac., San Francisco, p. 153 

\bibitem[\protect\citeauthoryear{Lehmann et al.}{2013}]{lehmann2013}
Lehmann H., Southworth J., Tkachenko A., Pavlovski K., 2013, A\&A, 557, A79 

\bibitem[{Lenz \& Breger(2004)}]{lenz2004}
Lenz, P. \& Breger, M. 2004, in {IAU} Symposium, Vol. 224, The {A-Star} Puzzle,
  786--790

\bibitem[{Liakos {et~al.}(2012)Liakos, Niarchos, Soydugan, \&
  Zasche}]{liakos2012}
Liakos, A., Niarchos, P., Soydugan, E., \& Zasche, P.\ 2012, MNRAS, 422, 1250

\bibitem[\protect\citeauthoryear{Maceroni et 
al.}{2014}]{maceroni2014} Maceroni C., et al., 2014, A\&A, 563, A59 

\bibitem[\protect\citeauthoryear{Malkov}{2003}]{malkov2003} Malkov O.~Y., 2003, A\&A, 402, 1055

\bibitem[{Malkov {et~al.}(2006)Malkov, Oblak, Snegireva, \& Torra}]{malkov2006}
Malkov, O.~Y., Oblak, E., Snegireva, E.~A., \& Torra, J.\ 2006, A\&A, 446, 785

\bibitem[{{Mas-Hesse} {et~al.}(2003){Mas-Hesse}, Gim\'{e}nez, Culhane, Jamar,
  {McBreen}, Torra, Hudec, Fabregat, Meurs, Swings, Alcacera, Balado,
  Beiztegui, Belenguer, Bradley, Caballero, Cabo, Defise, D\'{i}az, Domingo,
  Figueras, Figueroa, Hanlon, Hroch, Hudcova, Garc\'{i}a, Jordan, Jordi,
  Kretschmar, Laviada, March, Mart\'{i}n, Mazy, Men\'{e}ndez, Mi, de~Miguel,
  Mu\~{n}oz, Nolan, Olmedo, Plesseria, Polcar, Reina, Renotte, Rochus,
  S\'{a}nchez, San~Mart\'{i}n, Smith, Soldan, Thomas, Tim\'{o}n, \&
  Walton}]{mas-hesse2003}
{Mas-Hesse}, J.~M., Gim\'{e}nez, A., Culhane, J.~L., {et~al.}\ 2003, A\&A, 411,
  L261

\bibitem[{Mason {et~al.}(2001)Mason, Gies, \& Hartkopf}]{mason2001}
Mason, B.~D., Gies, D.~R., \& Hartkopf, W.~I.\ 2001, ASSL, 264, 37

\bibitem[{Mason et al.(2009)}]{mason2009} 
Mason, B.~D., Hartkopf, W.~I., Gies, D.~R., Henry, T.~J., \& Helsel, J.~W. 2009, AJ, 137,
3358 

\bibitem[{Mayor {et~al.}(2001)Mayor, Udry, Halbwachs, \& Arenou}]{mayor2001}
Mayor, M., Udry, S., Halbwachs, J., \& Arenou, F.\ 2001, in {IAU} Symposium,
  Vol. 200, The Formation of Binary Stars, Postdam, Germany, p. 45

\bibitem[{{McLaughlin}(1924)}]{mclaughlin1924}
{McLaughlin}, D.~B. 1924, {ApJ}, 60, 22

\bibitem[{Michalska \& Pigulski(2007)}]{michalska2007}
Michalska, G. \& Pigulski, A.\ 2007, Communications in Asteroseismology, 150, 71

\bibitem[{Mihalas(1978)}]{mihalas1978}
Mihalas, D.\ 1978, Stellar Atmospheres (W. H. Freeman)

\bibitem[{Mkrtichian {et~al.}(2007)Mkrtichian, Kim, Rodr\'{i}guez, Olson,
  Nazarenko, Gamarova, Kusakin, Lehmann, Lee, \& Kang}]{mkrtichian2007}
Mkrtichian, D.~E., Kim, S., Rodr\'{i}guez, E., {et~al.}\ 2007, in O. Demircan, S. O. Selam, and B. Albayrak, eds, ASP Conf. Ser., Vol. 370, Solar and Stellar Physics Through Eclipses. Astron. Soc. Pac., San Francisco, p. 194

\bibitem[{Mkrtichian {et~al.}(2004)Mkrtichian, Kusakin, Rodriguez, Gamarova,
  Kim, Kim, Lee, Youn, Kang, Olson, \& Grankin}]{mkrtichian2004}
Mkrtichian, D.~E., Kusakin, A.~V., Rodriguez, E., {et~al.}\ 2004, A\&A, 419,
  1015

\bibitem[{Mkrtichian {et~al.}(2003)Mkrtichian, Nazarenko, Gamarova, Lehmann,
  Rodriguez, Olson, Kim, Kusakin, \& {Rovithis-Livaniou}}]{mkrtichian2003}
Mkrtichian, D.~E., Nazarenko, V., Gamarova, A.~Y., {et~al.}\ 2003, in Interplay
  of Periodic, Cyclic and Stochastic Variability in Selected Areas of the {H-R}
  Diagram, Vol. 292, 113

\bibitem[{Paschke \& Brat(2006)}]{paschke2006}
Paschke, A. \& Brat, L.\ 2006, Open European Journal on Variable Stars, 23, 13

\bibitem[{Pigulski \& Michalska(2007)}]{pigulski2007}
Pigulski, A. \& Michalska, G.\ 2007, Acta Astron., 57, 61

\bibitem[{Pr\v{s}a \& Zwitter(2005)}]{pra2005}
Pr\v{s}a, A. \& Zwitter, T.\ 2005, ApJ, 628, 426

\bibitem[{Pr\v{s}a {et~al.}(2008)Pr\v{s}a, Guinan, Devinney, \&
  Engle}]{pra2008}
Pr\v{s}a, A., Guinan, E.~F., Devinney, E.~J., \& Engle, S.~G.\ 2008, A\&A, 489,
  1209

\bibitem[{Pr\v{s}a {et~al.}(2011)Pr\v{s}a, Batalha, Slawson, Doyle, Welsh,
  Orosz, Seager, Rucker, Mjaseth, Engle, Conroy, Jenkins, Caldwell, Koch, \&
  Borucki}]{pra2011}
Pr\v{s}a, A., Batalha, N., Slawson, R.~W., {et~al.}\ 2011, AJ, 141, 83

\bibitem[{Raskin {et~al.}(2011)Raskin, van Winckel, Hensberge, Jorissen,
  Lehmann, Waelkens, Avila, de~Cuyper, Degroote, Dubosson, Dumortier,
  Fr\'{e}mat, Laux, Michaud, Morren, Perez~Padilla, Pessemier, Prins, Smolders,
  van Eck, \& Winkler}]{raskin2011}
Raskin, G., van Winckel, H., Hensberge, H., {et~al.}\ 2011, A\&A, 526,

\bibitem[{Reegen(2007)}]{reegen2007}
Reegen, P.\ 2007, A\&A, 467, 1353

\bibitem[{Rodr\'{i}guez \& Breger(2001)}]{rodriguez2001}
Rodr\'{i}guez, E. \& Breger, M.\ 2001, A\&A, 366, 178

\bibitem[{Rodr\'{i}guez {et~al.}(2007)Rodr\'{i}guez, Garc\'{i}a, Costa, van
  Cauteren, Lampens, Olson, Amado, {L\'{o}pez-Gonz\'{a}lez}, Rolland,
  L\'{o}pez~de Coca, Turcu, Kim, Zhou, Wood, Hintz, Pop, Moldovan, Etzel, Lee,
  Handler, \& Mkrtichian}]{rodriguez2007}
Rodr\'{i}guez, E., Garc\'{i}a, J.~M., Costa, V., {et~al.}\ 2007, Communications
  in Asteroseismology, 150, 63

\bibitem[{Rodr\'{i}guez {et~al.}(2010)Rodr\'{i}guez, Garc\'{i}a, Costa,
  Lampens, van Cauteren, Mkrtichian, Olson, Amado, {Daszy\'nska-Daszkiewicz},
  Turcu, Kim, Zhou, {L\'{o}pez-Gonz\'{a}lez}, Rolland, {D\'{i}az-Fraile}, Wood,
  Hintz, Pop, Moldovan, Etzel, Casanova, Sota, Aceituno, \&
  Lee}]{rodriguez2010}
Rodr\'{i}guez, E., Garc\'{i}a, J.~M., Costa, V., {et~al.}\ 2010, MNRAS, 408,
  2149

\bibitem[{Rossiter(1924)}]{rossiter1924}
Rossiter, R.~A.\ 1924, ApJ, 60, 15

\bibitem[{Rucinski(1969)}]{rucinski1969}
Rucinski, S.~M. 1969, Acta Astron., 19, 245

\bibitem[Samus et al.(2009)]{samus2009} 
Samus, N.~N., Durlevich, O.~V., \& et al.\ 2009, VizieR Online Data Catalog, 1, 2025 

\bibitem[Sana 
\& Evans(2011)]{sana2011} Sana, H., \& Evans, C.~J.\ 2011, IAU Symposium, 272, 474 

\bibitem[Schmidt-Kaler (1982)]{schmidt-kaler1982} Schmidt-Kaler, T.,\ 1982, in: Landolt-B\"ornstein, Vol. 2b, eds.\ K. Schaifers, H. H. Voig. Springer, Heidelberg

\bibitem[{Soydugan {et~al.}(2006)Soydugan, Soydugan, Demircan, \&
  \.{I}bano\v{g}lu}]{soydugan2006}
Soydugan, E., Soydugan, F., Demircan, O., \& \.{I}bano\v{g}lu, C. 2006, MNRAS, 370, 2013

\bibitem[{Soydugan {et~al.}(2009)Soydugan, Soydugan, Senyuz, Puskullu, Tuysuz,
  Bakis, Bilir, \& Demircan}]{soydugan2009}
Soydugan, E., Soydugan, F., Senyuz, T., {et~al.}\ 2009, IBVS, 5902, 1

\bibitem[{Soydugan {et~al.}(2010)Soydugan, Soydugan, \c{S}eny\"{u}z,
  T\"{u}ys\"{u}z, Baki\c{s}, Bilir, \c{C}i\c{c}ek, \& Demircan}]{soydugan2010}
Soydugan, E., Soydugan, F., \c{S}eny\"{u}z, T., {et~al.}\ 2010, ASP Conf. Ser. Vol. 435, Binaries - Key to Comprehension of the Universe. Astron. Soc. Pac., San Francisco, p. 331
  
\bibitem[{Soydugan {et~al.}(2011)Soydugan, Soydugan, \c{S}eny\"{u}z,
  P\"{u}sk\"{u}ll\"{u}, \& Demircan}]{soydugan2011}
Soydugan, E., Soydugan, F., \c{S}eny\"{u}z, T., P\"{u}sk\"{u}ll\"{u}, c., \&
  Demircan, O.\ 2011, New A, 16, 72

\bibitem[{Stellingwerf(1978)}]{Stellingwerf1978}
Stellingwerf, R.~F.\ 1978, ApJ, 224, 953

\bibitem[{Svechnikov \& Kuznetsova(1990)}]{svechnikov1990}
Svechnikov, M.~A. \& Kuznetsova, E.~F.\ 1990, Katalog priblizhennykh
  fotometricheskikh i absoliutnykh elementov zatmennykh peremennykh zvezd:
  Izd-vo Ural'skogo universiteta, 1990- (Sverdlovsk : Izd-vo Ural'skogo
  universiteta, 1990-)

\bibitem[{Walker {et~al.}(2003)Walker, Matthews, Kuschnig, Johnson, Rucinski,
  Pazder, Burley, Walker, Skaret, Zee, Grocott, Carroll, Sinclair, Sturgeon, \&
  Harron}]{walker2003}
Walker, G., Matthews, J., Kuschnig, R., {et~al.}\ 2003, PASP, 115, 1023

\bibitem[{Watson {et~al.}(2012)Watson, Henden, \& Price}]{watson2012}
Watson, C., Henden, A.~A., \& Price, A.\ 2012, {VizieR} Online Data Catalog, 1,
  02027

\bibitem[{Watson(2006)}]{watson2006}
Watson, C.~L.\ 2006, Journal of the American Association of Variable Star
  Observers {(JAAVSO)}, 35, 318

\bibitem[{Wilson(1979)}]{wilson1979}
Wilson, R.~E.\ 1979, ApJ, 234, 1054

\bibitem[{Wilson \& Devinney(1971)}]{wilson1971}
Wilson, R.~E. \& Devinney, E.~J.\ 1971, ApJ, 166, 605

\bibitem[{Winkler {et~al.}(2003)Winkler, Courvoisier, Di~Cocco, Gehrels,
  Gim\'{e}nez, Grebenev, Hermsen, {Mas-Hesse}, Lebrun, Lund, Palumbo, Paul,
  Roques, Schnopper, Sch\"{o}nfelder, Sunyaev, Teegarden, Ubertini, Vedrenne,
  \& Dean}]{winkler2003}
Winkler, C., Courvoisier, T. J.~L., Di~Cocco, G., {et~al.}\ 2003, A\&A, 411, L1

\bibitem[{Zasche(2011)}]{zasche2011}
Zasche, P.\ 2011, New A, Volume 16, Issue 3, p. 157-160., 16

\bibitem[{Zhang {et~al.}(2013)Zhang, Zhou, Zhao, \& Wu}]{zhang2013}
Zhang, Y.~X., Zhou, X.~L., Zhao, Y.~H., \& Wu, X.~B. 2013,\ ApJ, 145, 42

\bibitem[{Zhou(2010)}]{zhou2010}
Zhou, A.~Y.\ 2010, {ArXiv} e-prints, 1002, 2729

\end{thebibliography}
\end{document}